\documentclass{PoS}

\title{Moments of net-charge multiplicity distribution in Au+Au collisions measured by the PHENIX experiment at RHIC}

\ShortTitle{Moments of net-charge multiplicity distribution  in Au+Au collisions ....}

\author{\speaker{P. Garg } (for PHENIX collaboration)\\
        Department of Physics, Banaras Hindu University, Varanasi-221005, India\\
        E-mail: \email{prakhar@rcf.rhic.bnl.gov}}

\abstract{
Beam Energy Scan (BES) program at RHIC is important to search for the existence of the critical point in the QCD phase diagram. Lattice QCD have shown that the predictions of the susceptibility of the medium formed in heavy-ion collisions can be sensitive to the various moments (mean ($\mu$) =${<x>}$, variance ($\sigma^2$) = ${<(x-\mu)^2>}$, skewness (S) = $\frac{<(x-\mu)^3>}{\sigma^3}$ and kurtosis ($\kappa$) =$\frac{<(x-\mu)^4>}{\sigma^4}  -3$ ) of conserved quantities like net-baryon number ($\Delta$B), net-electric charge ($\Delta$Q) and net-strangeness ($\Delta$S).
 Any non-monotonic behavior of the higher moments would confirm the existence of the QCD critical point. The recent results of the higher moments of net-charge multiplicity distributions for Au+Au collisions at $\sqrt{s}_{NN}$ varying from 7.7 GeV to 200 GeV from the PHENIX experiment at RHIC are presented.  
The energy and centrality dependence of the higher moments and their products (S$\sigma$ and $\kappa\sigma^{2}$) are shown for the net-charge multiplicity distributions. Furthermore, the results are compared with the values obtained from the heavy-ion collision models, where there is no QCD phase transition and critical point.
}

\FullConference{$8^{th}$ International Workshop on Critical Point and Onset of Deconfinement\\
		March 11 - 15, 2013\\
		Napa, California, USA}

\begin{document}

\section{Introduction}
The phenomenology of Quantum Cromodynamics (QCD) at finite temperature and baryon number density is one of the least explored regimes of the theory \cite{Stephanov}. QCD predicts a phase transition from hadron gas (HG) phase to quark gluon plasma (QGP) phase at high temperature and/or baryon density. The exact nature of the phase transition is still not established, however, various QCD based model indicate that at large $\mu_{B}$ and lower T the transition from hadronic phase to the Quark Gluon-Plasma (QGP) phase is of first order. On the other hand, lattice QCD calculations with physical quark masses suggests that the phase transition at high T and lower $\mu_{B}$ could be a simple cross over from QGP to hadron phase. It further suggest that the first order phase transition line should end somewhere at finite $\mu_{B}$ and T and that point will be a critical point of second order\cite{Stephanov}. 

It has been proposed that a critical point is a genuine thermodynamic singularity at which susceptibilities diverge and the order parameter fluctuates on long wavelengths \cite{Stephanov:1999zu}. But all the signatures share the common property that they are non-monotonic as a function of an experimentally varied parameter such as collision energy \cite{Stephanov:2008qz}. Typically, most of the fluctuation measures are related to quadratic variances of event-by-event observables, such as particle multiplicities, net charge, baryon number, particle ratios or mean transverse momentum in the event. However, higher non-Gaussian moments of the fluctuations are much more sensitive to the proximity of the critical point than the commonly employed measures based on quadratic moments\cite{Stephanov:2008qz}. Further, It has been suggested by lattice QCD calculations that the first three cumulants of net electric charge fluctuations are well suited for a determination of freeze-out parameters in a heavy ion collision \cite{Bazavov:2012vg}.

Recently, Beam Energy Scan (BES) program at relativistic heavy ion collider (RHIC) has been started to search the location of critical point. It has been suggested that the medium created in heavy ion collision experiments at different center of mass energy ($\sqrt{s_{NN}}$) follow different trajectories on the temperature (T) and baryon chemical potential ($\mu_{B}$) plane during their time evolution \cite{Stephanov}. This will enable us to  locate the critical point in T-$\mu_{B}$ plane by experimentally measuring the fluctuations in higher moments of net charge, net baryon etc. with respect to $\sqrt{s_{NN}}$. In the present work, fluctuations of the net charge and it's higher moments are obtained from the net charge distributions measured by PHENIX experiment at RHIC. \\

\section {Analysis details}
In the present work, 200 GeV Au+Au data of RUN07 and 62.4 GeV, 39 GeV and 7.7 GeV data of RHIC RUN10 taken by PHENIX experiment is used. The event-by-event net-charge distribution is measured for Au+Au collisions occurring within $\pm$30 cm along the z position of the interaction point. The charged particles are measured between the transverse momentum ($p_{T}$) range $0.3~GeV/c < p_{T} < 1.0 GeV/c$ and pseudorapidity ($\eta$) range at $|\eta|$ < 0.35 region. The standard PHENIX track quality cuts are used for this analysis. To avoid the auto-correlation effect in the higher moments analysis,  the centrality selection has been done by using tracks in a different pseudo rapidity range. The finite centrality bin width effect has been corrected by using a centrality bin width correction. Delta theorem is used  for the statistical error estimation of higher moments \cite{luo:luo}. \\

\begin{figure}
\centering
\includegraphics[width=0.5\textwidth]{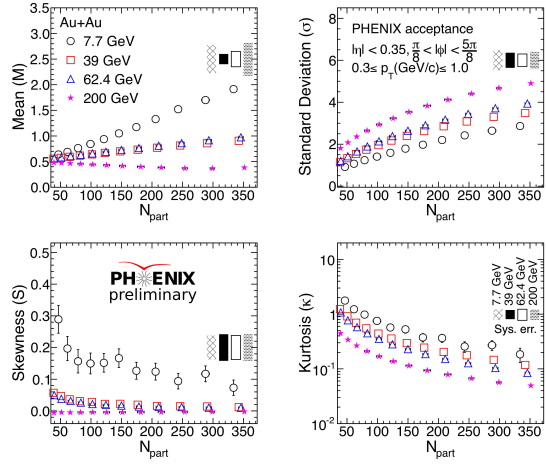}
\caption{(Color online) Mean, Standard Deviation ($\sigma$), Skewness and Kurtosis obtained from net-charge distributions for Au+Au collision at $\sqrt{s_{NN}}$ = 200 GeV, 62.4 GeV, 39 GeV and 7.7 GeV  for most-central to most-peripheral events are shown.}
\label{fig:dist}
\end{figure}

\begin{figure}[!h]
  \begin{center}
    \includegraphics[width=58mm]{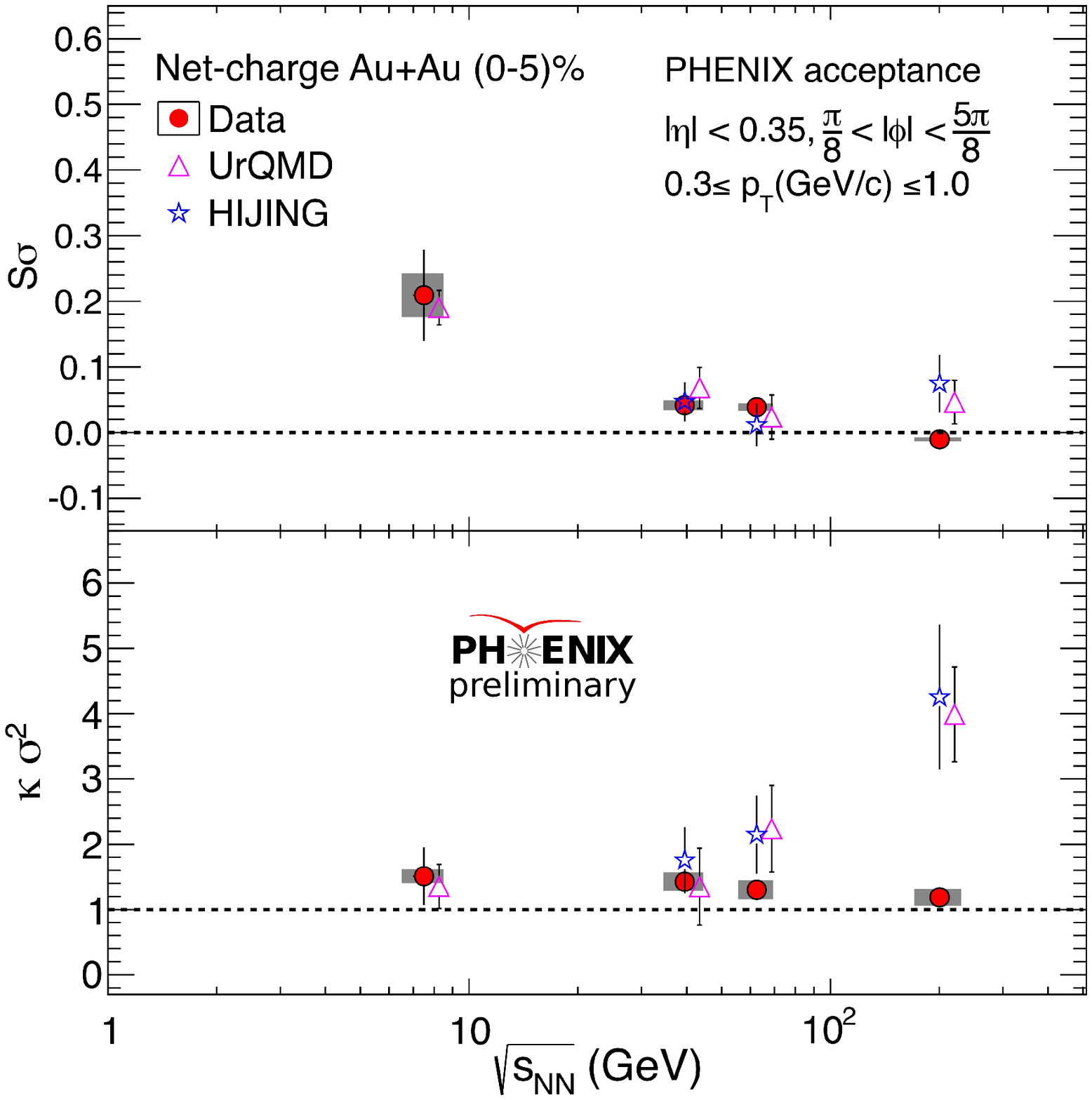}
    \hspace{5mm}
    \includegraphics[width=58mm]{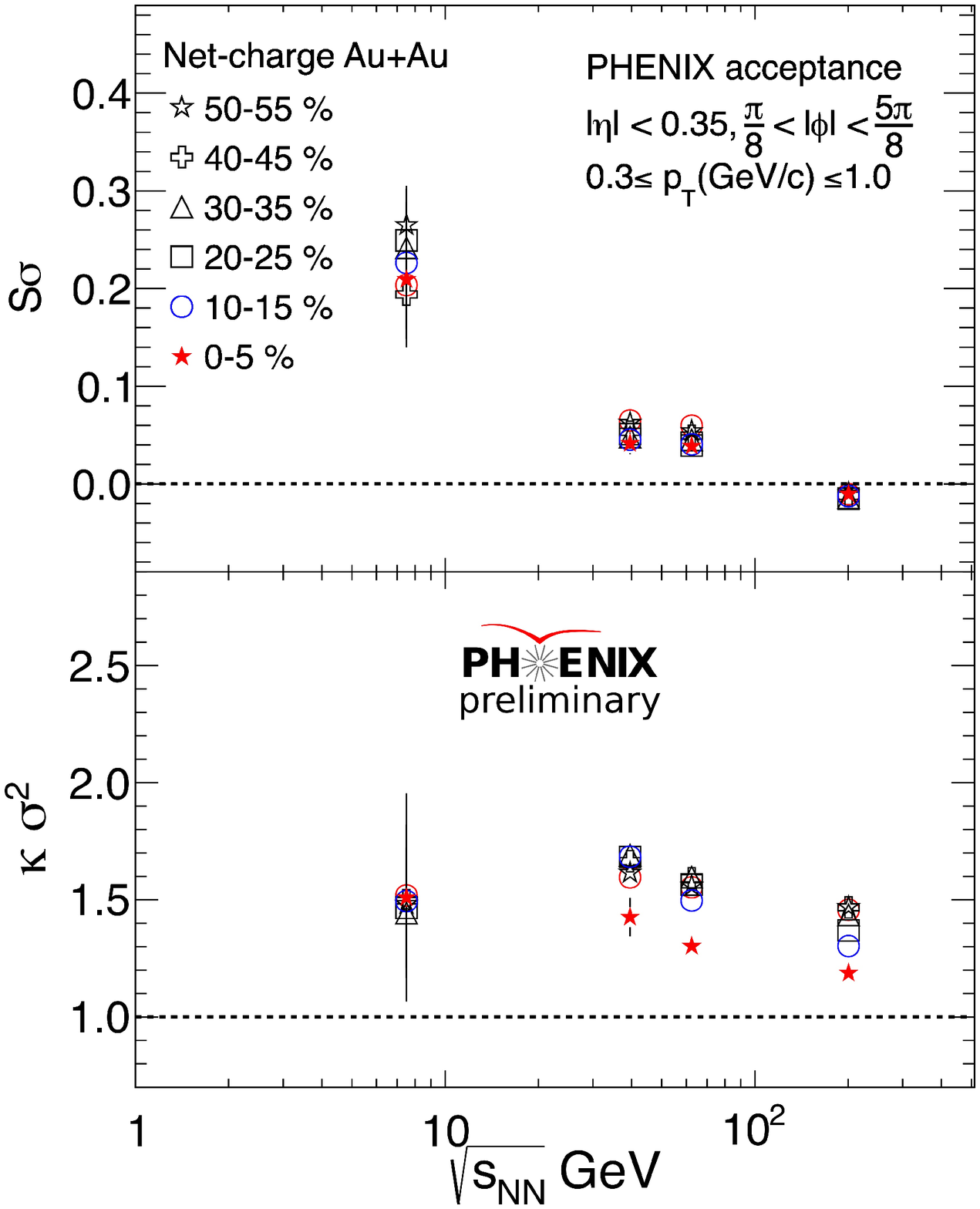} \newline%
   (a) \hspace{80mm} (b)
  \end{center}
  \begin{quote}
    \caption[]{(Color online) The skewness multiplied by the standard deviation and the kurtosis multiplied by the variance from net charge distributions for Au+Au collisions as a function of $\sqrt{s_{NN}}$. (a) The circles represent the data for most central collision events. The grey error bars represent the systematic errors. (b) $S\sigma$ and $\kappa\sigma^{2}$ for different centralities as a function of $\sqrt{s_{NN}}$.
 \label{fig:excite}} %
  \end{quote} %
\end{figure} 
\vspace{0 cm}
\section{Results}
Mean, Standard Deviation, Skewness and Kurtosis, calculated from event-by-event net-charge multiplicity distributions for Au+Au collisions at $\sqrt{s_{NN}}$ = 7.7 GeV, 39 GeV, 62.4 GeV and 200 GeV are shown in Fig. \ref{fig:dist} as a function of number of participants ($N_{part}$). Mean and Standard deviation increases with $N_{part}$ while Skewness and Kurtosis decreases with $N_{part}$, although for 200 GeV, mean is almost constant within the systematic errors. Also the Mean, Skewness, and Kurtosis increases with decreasing $\sqrt{s_{NN}}$ while the trend is opposite in case of Standard Deviation as is evident from Fig. \ref{fig:dist}. 

The skewness and kurtosis, calculated from the net-charge distributions,  are expressed in terms that can be associated with the quark number susceptibilities, $\chi: S \sigma \approx \chi(3)/\chi(2) $ and $\kappa\sigma^2 \approx \chi(4)/\chi(2)$\cite{Karsch}. Theoretically, if we take the ratio of the susceptibilities, the volume effect is canceled out. The energy dependence of  $S\sigma$ and $\kappa\sigma^2$ of net-charge distributions are shown in Fig. \ref{fig:excite}. The statistical and systematic uncertainties are shown along with the data points. The experimental values are compared with the model calculations from UrQMD and HIJING within the PHENIX acceptance.

The HIJING and URQMD results match the data points except at highest energies, which may be due to the contribution of resonance production. $S\sigma$ and $\kappa\sigma^2$ are shown for different centralities as a function of $\sqrt {s_{NN}}$ as shown in Fig.  \ref{fig:excite} (b).

\section {Summary}
The higher moments of the net-charge multiplicity distributions have been measured for Au+Au collisions at $\sqrt{s_{NN}}$ = 7.7 to 200 GeV. The values of $S\sigma$ decreases with $\sqrt{s_{NN}}$ while $\kappa\sigma^2$ remains constant within uncertainties for all centralities. There is no significant deviation from the simulation results as observed in the data at these four collision energies. It will be interesting to study at lower energies like 19.6 GeV and 27 GeV which are still under investigation.

\section {Acknowledgement}
PG acknowledges the financial support from CSIR, New Delhi, India.  

 \end{document}